\documentclass{aa}  

\usepackage{graphicx}

\usepackage{natbib}
\usepackage{amsmath}

\newcommand{\logg}{\ensuremath{\log g}}

\newcommand{\teff}{$T_{\rm eff}$}
\newcommand{\msun}{$M_{\odot}$}
\newcommand{\kms}{${\rm km\;s^{-1}}$}

\usepackage{txfonts}
\begin{document}

\title{The $^6$Li/$^7$Li isotopic ratio in the metal-poor binary CS22876--032
\thanks{Based on observations made with the Very Large Telescope (VLT) at 
ESO Paranal Observatory, Chile, Programme 080.D-0333}}


\author{J.~I. Gonz\'alez Hern\'andez
        \inst{1,2}
        \and
        P. Bonifacio\inst{3}
        \and
        E. Caffau\inst{3}
        \and
        H.-G. Ludwig\inst{4}
        \and
        M. Steffen\inst{5}
        \and
        L. Monaco\inst{6}
        \and
        R. Cayrel\inst{3}
        }

\institute{Instituto de Astrof\'{\i}sica de Canarias, 
V\'{\i}a L\'actea, 38205 La Laguna, Tenerife, Spain\\
\email{jonay@iac.es}
        \and
Universidad de La Laguna, Departamento de Astrof\'{\i}sica, 
38206 La Laguna, Tenerife, Spain
        \and
GEPI, Observatoire de Paris, Universit{\'e} PSL, CNRS, 5 Place Jules 
Janssen, 92190 Meudon, France
        \and
Zentrum f{\"u}r Astronomie der Universit{\"a}t Heidelberg, 
Landessternwarte, K{\"o}nigstuhl 12, 69117 Heidelberg, Germany
        \and
Leibniz-Institut f{\"u}r Astrophysik Potsdam (AIP), An der Sternwarte 16, 
14482 Potsdam, Germany
        \and
Departamento de Ciencias Fisicas, Universidad Andres Bello, Fernandez 
Concha 700, Las Condes, Santiago, Chile        
        }

   \date{Received June 03, 2019; accepted July xx, 2019}

 
  \abstract
  {}
{We present high-resolution and high-quality UVES spectroscopic data of the 
metal-poor double-lined spectroscopic binary CS 22876--032 ([Fe/H]~$\sim -3.7$~dex). Our  goal is to derive the $^6$Li/$^7$Li isotopic ratio by analysing the \ion{Li}{i} 
$\lambda$~670.8~nm doublet.}
{We co-added all 28 useful spectra normalised and corrected for radial 
velocity to the rest frame of the primary star. 
We fitted the Li profile with a grid of the 3D non-local thermodynamic equilibrium (NLTE) synthetic spectra 
to take into account the line profile asymmetries induced by stellar convection, 
and performed Monte Carlo simulations to evaluate the uncertainty of the fit of the Li 
line profile.}
{We checked that the veiling factor does not affect the derived isotopic ratio,
$^6$Li/$^7$Li, and only modifies the Li abundance, A(Li), by about 0.15~dex. 
The best fit of the Li profile of the primary star provides 
A(Li)~$ = 2.17 \pm 0.01$~dex and $^6$Li/$^7$Li~$=8^{+2}_{-5}$\% at 68\% 
confidence level.
In addition, we improved the Li abundance of the secondary star at 
A(Li)~$= 1.55 \pm 0.04$~dex, which is about 0.6~dex lower than that of the 
primary star.}
{The analysis of the Li profile of the primary star is consistent with 
no detection of $^6$Li and provides an upper limit to the isotopic ratio 
of $^6$Li/$^7$Li~$< 10$\% at this very low metallicity, about 0.5~dex lower 
in metallicity than previous attempts for detection of $^6$Li in extremely metal 
poor stars.
These results do not solve or worsen the cosmological $^7$Li 
problem, nor do they support the need for non-standard $^6$Li production 
in the early Universe.}


   \maketitle
%

\section{Introduction} \label{sec:intro}

The standard Big Bang nucleosynthesis (SBBN) theory predicts a very small primordial 
production of lithium in the first minutes of the Universe, at about ten orders of 
magnitude lower than H and He \citep[e.g.][]{ste07arnps,cyb16rmp}. This prediction 
mainly depends on the density of baryons inferred from the fluctuations of the
cosmic microwave background (CMB) as measured by the Wilkinson Microwave Anisotropy Probe (WMAP)~\citep{spe07apjs} and  the 
Planck satellite~\citep{planck16aa}.
This   small amount of Li predicted, ranging from 
A($^7$Li)\footnote{$A({\rm X})=\log[N({\rm X})/N({\rm H})]+12$}~$=2.67$~\citep{cyb16rmp} 
to 2.74 \citep{coc17},
is still   a factor of three greater than the almost uniform measured Li 
content in the atmosphere 
of unevolved metal-poor field stars \citep[e.g.][]{spi82nat,reb88aa,bon07aaI,aok09apj} 
or in globular clusters~\citep[e.g.][]{bon07aaII,gon09aa}. 

A variety of different models have been 
proposed to try to explain this abundance difference known as the cosmological lithium 
problem~\citep[see e.g.][and references therein]{fie14bbn}. 
However, more recent studies have shown a decline and/or an increasing scatter in 
the Li abundances in metal-poor stars for metallicities [Fe/H]~$<-3$
\citep[e.g.][]{sbo10aa,bon12aa,mat17pasj}. In particular, the flame of the cosmological Li problem 
has  recently been relighted with the Li detection in some extremely iron-poor 
stars~\citep{han14apj,sta18mnras,bon18aa,agu19apjl}, in contrast with the Li 
non-detections in apparently similar stars~\citep{fre05nat,caf11nat,bon15aa}. 
Two extremely iron-poor stars, SDSS\,J0135+0641~\citep{bon15aa,bon18aa} at 
[Fe/H]~$< -5.2$, and SDSS\,J0023+0307~\citep{agu18apjlII,agu19apjl} at [Fe/H]~$< -6.1$, 
almost recover  the level of the lithium plateau, with lithium abundances of 
A(Li) $\sim 1.9$ and $\sim2.0$, respectively. 
These recent discoveries pose strong constraints on any 
theory aiming to explain the cosmological Li problem.
  
The SBBN prediction for the $^6$Li isotope is even lower, at about 
A($^6$Li)~$\sim -1.9$~\citep{cyb16rmp}, 
i.e. a tiny isotopic ratio of ${\rm {^6}Li}/{\rm {^7}Li} = 2.75 \times 10^{-5}$, not 
measurable through spectral analysis of stars. 
However, $^6$Li, like the B and Be isotopes, can be efficiently produced by the 
interaction of energetic nuclei of Galactic cosmic rays (GCR) with the nuclei of the 
interstellar medium (ISM), where the $^6$Li abundance should 
increase as the metallicity increases~\citep{pra12aa}. The predicted level of 
the isotopic ratio ${\rm {^6}Li}/{\rm {^7}Li}$ is 1\% at metallicity [Fe/H]~$=-2$.
Detecting higher amounts of $^6$Li at low metallicities may suggest other production 
channels, for example  non-standard physics in the Big Bang~\citep{jed09njph}, 
or a pre-galactic origin~\citep{rol06apj}.

The presence of $^6$Li in metal-poor halo stars can only be derived from the 
asymmetry in the red wing of the $^7$Li doublet line at $\lambda$~670.8~nm. The 
treatment of this line using 1D model atmospheres and LTE spectral synthesis 
produced presumably the detection of $^6$Li in several stars at 
[Fe/H]~$< -2$~\citep{cay99aa,asp06apj}.
However, convective flows in the atmospheres of metal-poor stars are likely  
responsible for part of this asymmetry~\citep{cay07aa}.
The re-analysis of the Li feature, using 3D hydrodynamical models and a 3D-NLTE 
treatment, in some metal-poor stars was not able to confirm the detection 
of $^6$Li~\citep{ste12msais,lin13aa}.

The metal-poor spectroscopic binary CS 22876--032, with very low metallicity at about 
[Fe/H]~$\sim -3.7$~\citep{mol90aa,nor00apj,gon08aa}, and its relatively large 
brightness ($m_V = 12.8$), provide a unique opportunity to search for $^6$Li  
in the primary star using high-resolution spectroscopic observations and 
3D-NLTE synthetic profiles. 
Here we report on the search for $^6$Li in CS 22876--032,  
and therefore a test performed at about 0.5~dex lower metallicity than other 
previous attempts to detect $^6$Li in extremely metal-poor stars. 
The Li abundances of both binary stellar components have been already reported, 
but only the primary star has a Li abundance at the level of the Spite 
plateau~\citep{gon08aa}. 

\begin{figure}
\begin{center}
{\includegraphics[clip=true,width=95mm,angle=0]{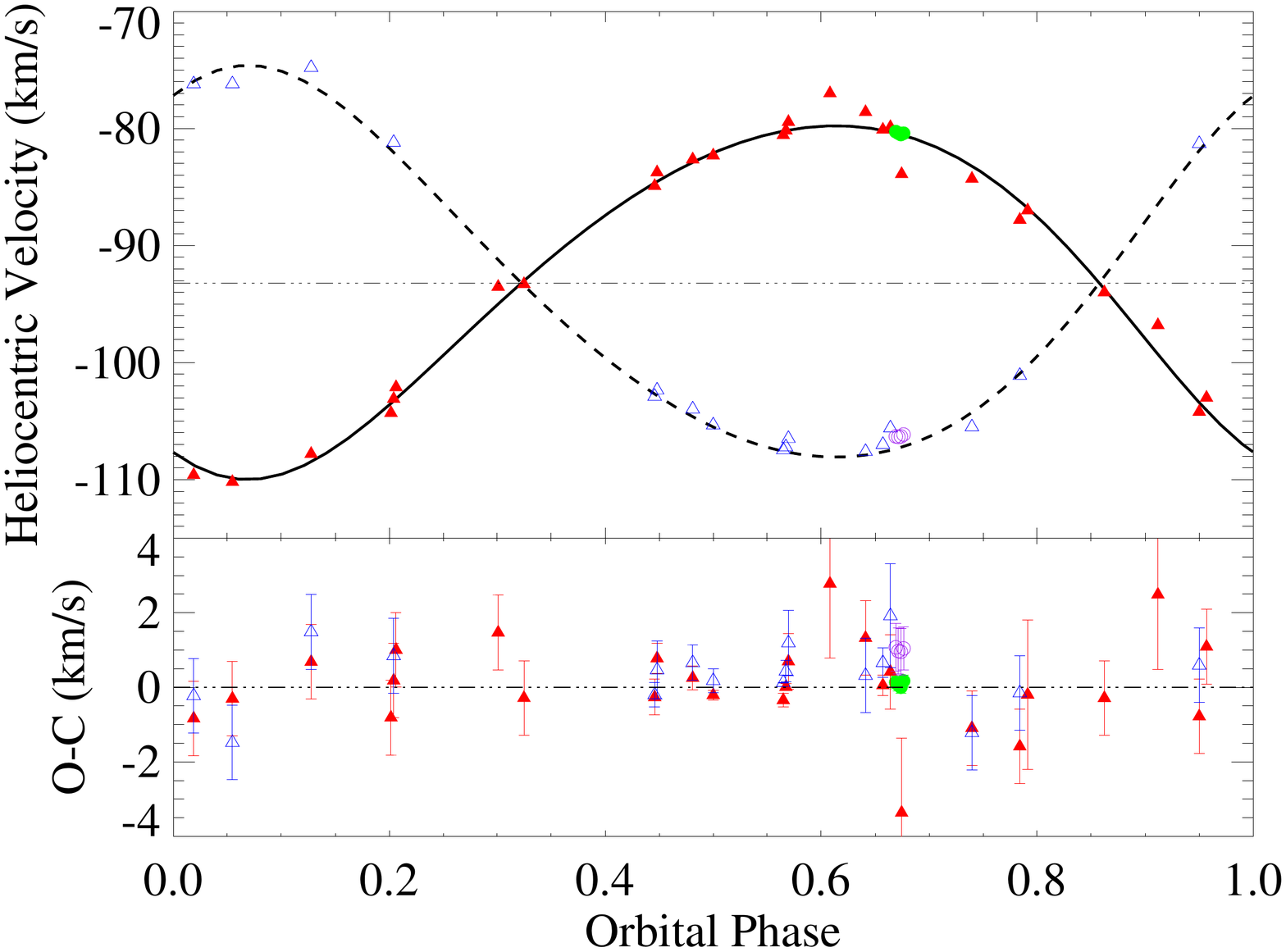}}
\end{center}
\caption{
Radial velocities of the primary 
(star A: filled symbols) and the secondary (star B: open symbols)
of CS 22876--032 (upper panel), and residuals from the fit (lower panel). 
Triangles and circles are literature~\citep{nor00apj,gon08aa} and new 
UVES observations, respectively.
The curves show the orbital solution (see Table~\ref{tab:rvc}) for star 
A (solid) and B (dashed). Triple dotted-dashed horizontal line shows the 
centre-of-mass velocity of the system.
\label{fig:rvc}
}
\end{figure}

\begin{figure}
\begin{center}
{\includegraphics[clip=true,width=95mm,angle=0]{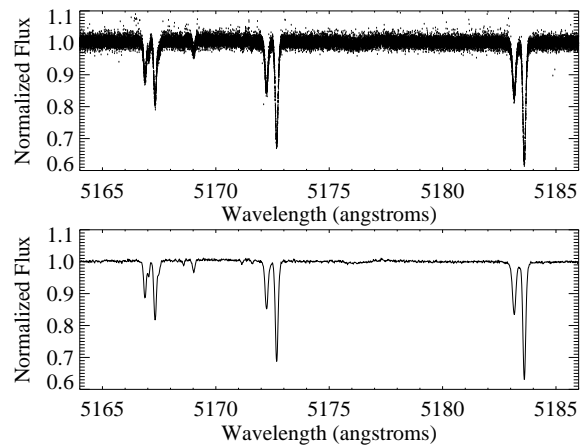}}
\end{center}
\caption{
Normalised, co-added, RV-corrected, unveiled spectrum
of CS 22876--032 shown in the spectral region of the \ion{Mg}{i}b triplet 
(upper panel) and rebinned to a wavelength step of 0.014 {\AA}/pixel 
(lower panel) in the rest frame of star A. 
\label{fig:spec}
}
\end{figure}

\section{Observations and data analysis} \label{sec:obs}

We carried out spectroscopic observations with UVES at UT2/VLT in Paranal (Chile) 
using the image slicer \#3 on 2007 October 17, 18, 19, and 20.
These dates were adequate since the orbital phase of the binary was $\sim 0.67$
where both stellar components were close to the maximum line-of-sight radial 
velocity difference (see Fig.~\ref{fig:rvc}). 
Another single observation with the same configuration was 
acquired on 2007 December 07 (with a median S/N$\sim 67$ per order), but was 
discarded in this analysis since it shows a different orbital phase with  
both stellar components closer in radial velocity. 
We obtained 28 useful observations of 3600s, covering 
the spectral region $\lambda\lambda$495.95--707.08~nm at a resolving power 
$R\sim107,200$. 
The reduced spectra were downloaded from the ESO Archive Science Portal. 
The median S/N per order is in the range 55-75, 
except for three spectra with a S/N$\sim 32$, 37, and 50 
with shorter exposure times of $\sim 1890$s, 2160s, and 2500s. 
The individual reduced order-merged spectra were first normalised in the whole 
wavelength range with a fifth-order polynomial using our own automated IDL-based 
routine 
and then corrected for barycentric velocity. 
Radial velocities (RV) were obtained by cross-correlating each spectrum 
with a mask centred at the \ion{Mg}{i}b triplet lines and modelled 
using a Gaussian with the full width half maximum (FWHM) equal to the resolving power 
(FWHM$=2.8$~\kms, $R\sim107,200$) of the observed spectrum and contrast equivalent
to relative depths of the \ion{Mg}{i}b triplet lines in the primary (star A).

\begin{table}[!ht]
\centering
\caption{Updated orbital elements of CS 22876--032\label{tab:rvc}}
\begin{tabular}{lrr}
\hline
\hline
Parameter & Value & Uncertainty\\
\hline
$P$ [days]                  &  425.15  &  0.13 \\
$T_0$ [HJD-2448500]         &  79.22   &  0.07 \\
$e$                         &  0.13    &  0.01 \\
$w$ [deg]                   &  147.8   &  0.4  \\
$\gamma$ [\kms]             &  -93.20  &  0.12 \\
$K_{\rm A}$ [\kms]          &  15.11   &  0.20 \\
$K_{\rm B}$ [\kms]          &  16.59   &  0.19 \\
$M_{\rm A}\sin^3 i$ [\msun] &  0.71  & 0.02 \\
$M_{\rm B}\sin^3 i$ [\msun] &  0.65 &  0.02 \\
$M_{\rm B}/M_{\rm A}$ [\msun]  &  0.91 &  0.02 \\
RMS${\rm (O-C)}_{\rm A}$ [\kms]  &  1.13 &  -- \\
RMS${\rm (O-C)}_{\rm B}$ [\kms]  &  0.87 &  -- \\
\hline
\end{tabular}
\end{table}

In Fig.~\ref{fig:rvc} we show the RV curve of the two stellar components,
with the RVs of the primary and the secondary at about $RV_A=-80.4\pm0.2$~\kms\ 
and $RV_B=-106.3\pm0.6$~\kms\ during those nights.
We compute the weighted RV average for each night and fit these RV points
together with those in \citet[][and references therein]{gon08aa} using 
the {\sc rvfit} tool~\citep{igl15pasp} to update 
the orbital parameters of the binary shown in Table~\ref{tab:rvc}. 

The individual spectra normalised to unity were corrected for the RV of star A 
and co-added all together. 
Only a small fraction of flux points, deviating more that 2$\sigma$ from 
the mean flux, were discarded.
In Fig.~\ref{fig:spec} we show a small portion, 
the \ion{Mg}{1}b triplet spectral region, of the normalised, co-added spectrum 
in the rest frame of the primary (star A). 
The S/N of this spectrum at continuum surrounding the \ion{Mg}{i}b 
triplet is S/N$\sim 100$. 
Finally, we rebinned the co-added spectrum with a wavelength step of 
0.014 {\AA}/pixel, by taking the weighted (by flux errors) mean of the 
flux points and respective wavelengths of each bin,
providing a S/N$\sim 570$ at the continuum (see Fig.~\ref{fig:spec}). 
The spectral lines of each binary component appear weakened because of the veiling
($f_A= 1.33$ and $f_B= 4.03$ at the \ion{Mg}{i}b triplet) that each star produces on 
the flux of the other component~\citep[see][for more details on the determination 
of veiling factors in this binary system]{gon08aa}.

\section{3D model atmospheres and non-LTE spectrum synthesis}

The lithium abundance and isotopic ratio are derived
by fitting synthetic line profiles of the \ion{Li}{i} $\lambda\,670.8$\,nm  
doublet to the observed spectrum. Synthetic spectra are
computed from 3D hydrodynamical CO5BOLD model atmospheres and account for
3D non-LTE effects.

The 3D model atmosphere chosen for the primary has a temporal average 
effective temperature $\left<T_{\rm eff}\right>=6550$\,K, surface gravity 
$\log g=4.5$, and metallicity [Fe/H]$=-4$. The atmosphere of the secondary is 
represented by a 3D model with $\left<T_{\rm eff}\right>=5920$\,K, 
$\log g=4.5$, and [Fe/H]$=-4$.
The non-local radiative transfer is solved in 12 opacity 
bins on a standard grid of $140\times 140\times 150$ cells
\citep[see][for further details on the 3D hydrodynamical model 
atmospheres]{lud09memsai}.
To reduce the burden of computing the synthetic line profiles, a number 
of representative snapshots were selected from the full time sequence of 
the two simulations (19 for the hotter model and 20 for the cooler).

As an intermediate step, the non-LTE departure coefficients 
$b_i=n_i({\rm NLTE})/n_i({\rm LTE})$ for each level $i$ of 
the 17 level lithium model atom with 34 line transitions were computed
as a function of geometrical position within all selected 3D model 
atmospheres  \citep[see e.g.][for details on the lithium model atom and the 
3D-NLTE treatment]{ste12msais,klevas16aa,mott17aa}. 
Since the Li resonance line is weak, the departure coefficients are 
essentially independent of the assumed
Li abundance within the considered range.

For both binary components, we then created a grid of 3D non-LTE synthetic 
spectra of the \ion{Li}{i} $\lambda\,670.8$\,nm doublet for different Li 
abundances and isotopic ratios. The synthetic line profiles are computed 
with the line formation code 
{\sc Linfor3D}\footnote{https://www.aip.de/Members/msteffen/linfor3d}, 
taking into account the detailed 3D thermal structure and hydrodynamical
velocity field of the selected snapshots and using the previously computed
NLTE departure coefficients.
 
\begin{figure}
\begin{center}
{\includegraphics[clip=true,width=85mm,angle=0]{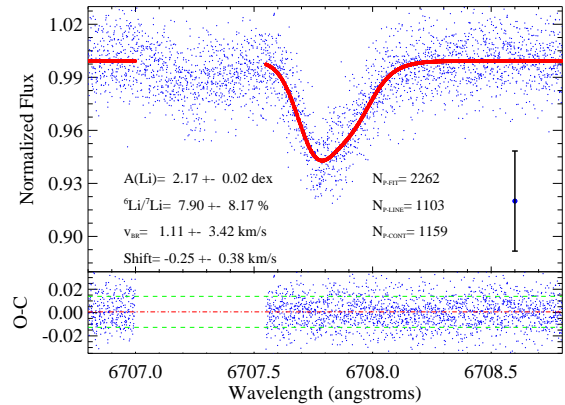}}
\end{center}
\caption{
Normalised, co-added, RV-corrected, unveiled spectrum 
of CS 22876--032A (blue dots), together with best-fit 3D-NLTE
Li profile (red circles), and residuals of the fit (lower panel). 
The mean flux uncertainty is shown on the right as a blue circle with error bars. The horizontal dashed and dot-dashed lines indicate the 1$\sigma$ dispersion 
of the observed flux points relative to the mean difference observed minus 
computed profiles.
\label{fig:lifit}
}
\end{figure}

\begin{figure}
\begin{center}
{\includegraphics[clip=true,width=90mm,angle=0]{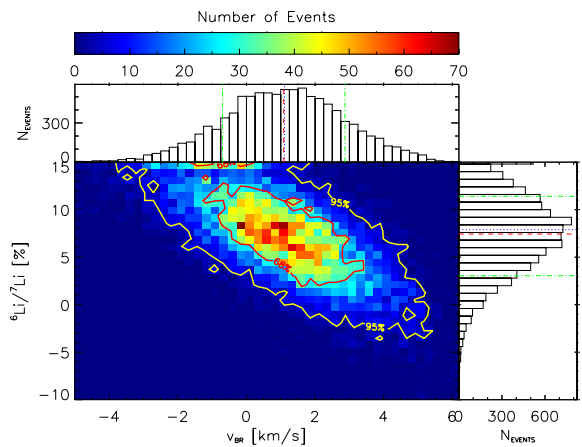}}
\end{center}
\caption{
Density distributions of lithium isotopic ratio vs. 
broadening velocity of 10,000 Monte Carlo simulations 
to evaluate the confidence of the best fit to the Li profile of star A. The green 
dash-dotted lines indicate the 1$\sigma$ uncertainty, whereas the blue dotted and 
red dashed lines are the best fit values and the centre of the distributions, 
respectively.
\label{fig:vbrmc}
}
\end{figure}

\section{Lithium profile of the primary star} \label{sec:pri}

The co-added spectrum in the rest frame of the primary star A is shown 
in Fig.~\ref{fig:lifit} for the spectral 
range $\lambda\lambda670.6-670.9$~nm. 
We used this spectrum without rebinning to measure 
the lithium isotopic ratio in the primary star. 
The large number of co-added spectra allowed us to perform the fit of the 
Li feature including about 1100 line flux points and 1160 continuum flux points  
in the spectral range $\lambda\lambda670.6-670.9$~nm.
The spectral lines of each binary component appeared weaker due to the corresponding 
veiling at the \ion{Li}{i} doublet spectral region. We corrected for the veiling at the 
\ion{Li}{1} doublet spectral region with the factors $f_A=1.36$ and $f_B=3.74$,
for the primary and secondary respectively~\citep{gon08aa}.

The grid was created for three different values of Li abundance 
(A(Li)$=$~1.6, 2.0 and 2.4), isotopic ratio ($^6$Li/$^7$Li$=$~0, 4 and 8 \%), 
and global broadening ($v_{\rm BR}=0$, 4.5 and 9 \kms), modelled as a Gaussian 
accounting for instrumental broadening, $v_{\rm INS}$, and rotational 
broadening, $v_{\rm ROT}$.
We performed a fit of the unbinned Li profile of the primary using an 
automated IDL-based routine that makes use of the 
{\sc MPFIT}\footnote{http://purl.com/net/mpfit} routine~\citep{mar09}, 
including five free parameters: 
continuum location, global Gaussian broadening, velocity shift, 
lithium abundance, and isotopic ratio. 
We allowed the fitting procedure to also explore  negative values of $v_{\rm BR}$, 
down to -5 km/s, by extrapolating in the grid of synthetic Li profiles.
The fitting procedure provides essentially the same $^6$Li/$^7$Li isotopic ratio 
for the unveiled and veiled spectrum of the primary star, but the Li abundance is lower  
by about 0.15 dex for the veiled spectrum. 
We therefore decided to run the fitting procedure on the unveiled spectrum.
In Fig.~\ref{fig:lifit} we show the best fit of the unbinned Li profile of star A, 
which gives $^6$Li/$^7$Li~$ = 7.9 \pm 8.2$\%, A(Li)~$ \sim 2.17\pm0.02$~dex, 
$v_{\rm BR}= 1.1 \pm 3.4$~\kms, and a global velocity shift of the line profile 
consistent with zero within the error bars. 
A S/N$\sim 74$ is inferred from the scatter of the 
fit residuals. 

\begin{figure}
\begin{center}
{\includegraphics[clip=true,width=90mm,angle=0]{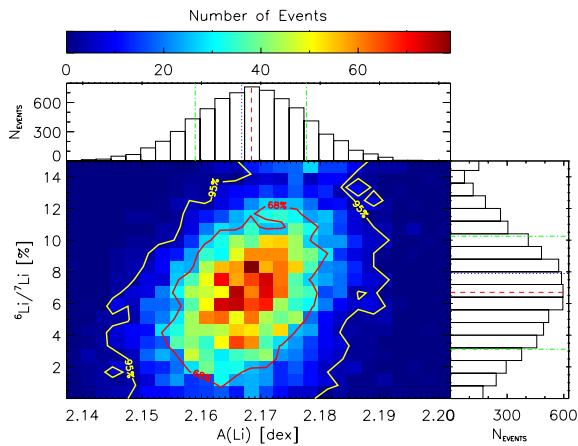}}
\end{center}
\caption{
Same as Fig.~\ref{fig:vbrmc}, but for the lithium isotopic ratio vs. 
lithium abundance. About 70\% of the simulations shown here 
fall at $v_{\rm BR} \ge 0$ and $^6$Li/$^7$Li~$\ge 0$.
\label{fig:limc}
}
\end{figure}

To evaluate the uncertainty on Li abundance and isotopic
ratio we ran 10,000 Monte Carlo simulations by injecting Poisson noise in the best fit 
according to the S/N at the continuum level of the unbinned spectrum.
We then repeated our fitting procedure 10,000 times, replacing the observed 
spectrum by each of the resulting 10,000 artificial realisations. 
In Fig.~\ref{fig:vbrmc} we show the 
distribution of the 10,000 Monte Carlo simulations of the lithium isotopic ratio versus 
the broadening velocity $v_{\rm BR}$. There is a clear correlation between
the isotopic ratio and the broadening velocity. The projection of these simulations
over broadening velocity shows a symmetric distribution and provides the same value
as the single best fit: $v_{\rm BR}= 1.1 \pm 1.8$~\kms. This value is lower than
the instrumental broadening ($v_{\rm INS}\sim2.8$~\kms), which points to a too broad CO5BOLD 
3D-NLTE line profile that needs to be further investigated. In Fig.~\ref{fig:limc} we 
show  the distributions of the Monte Carlo simulations for lithium isotopic ratio versus
the lithium abundance. 
About 30\% of the simulations fall at $v_{\rm BR}<0$~\kms and $^6$Li/$^7$Li~$<0$, 
which were discarded in Fig~\ref{fig:limc}.
Most of the remaining simulations fall at the same values of the initial 
best fit and the contours and density distributions of Li abundance and isotopic ratio
provide the same answer: $^6$Li/$^7$Li~$ = 7.9 ^{+2.3}_{-4.8}$\% and 
A(Li)~$ \sim 2.17\pm0.01$~dex. 
This defines an upper limit to the isotopic ratio of $^6$Li/$^7$Li~$<10.2$~\% 
and 13.8~\% at 1$\sigma$ and 2$\sigma$ confidence, respectively.

\begin{figure}
\begin{center}
{\includegraphics[clip=true,width=95mm,angle=0]{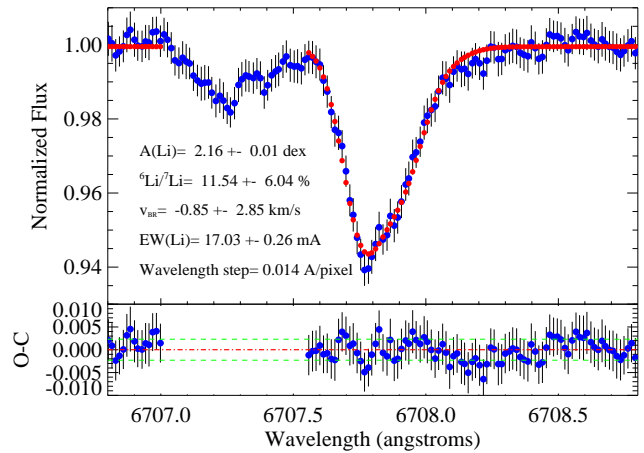}}
\end{center}
\caption{
Best fit (red circles) to the unveiled Li profile of star A (blue circles) 
rebinned to a wavelength step of 0.014 {\AA}/pixel (upper panel) 
and the residuals of the fit (lower panel). 
\label{fig:limfit}
}
\end{figure}

\begin{figure}
\begin{center}
{\includegraphics[clip=true,width=95mm,angle=0]{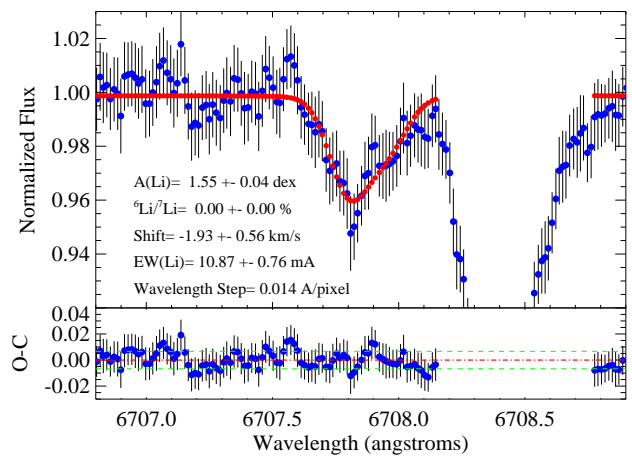}}
\end{center}
\caption{
Same as Fig.~\ref{fig:limfit}, but for star B.
\label{fig:limfitb}
}
\end{figure}

In Fig.~\ref{fig:limfit} we show the unveiled Li profile of the 
primary star rebinned with a wavelength step of 0.014~{\AA}/pixel. 
The fit of the rebinned spectrum provides a similar result, within the error bars, 
to that of the unbinned co-added spectrum, as expected~\citep{bon05msais}, 
but the result obtained using the unbinned spectrum is statistically more robust. 
The number of fitted points in the rebinned case are 50 and 54 for the line
and continuum, and we get a S/N$\sim$422 from the residuals of the fit. From
this best fit we measure the EW of the Li profile of the primary star at 
$17.0\pm0.3$~m{\AA}.

\section{Lithium profile of the secondary star} \label{sec:sec}

The RV corrected, rebinned, unveiled spectrum in the rest frame of the secondary  
is shown in Fig.~\ref{fig:limfitb}. The 
3D-NLTE grid was created for the A(Li)$=$~1.4, 1.8, and 2.2 
based on the 3D hydrodynamical model with 
$\left <T_{\rm eff}\right >=5920$\,K, $\log g=4.5$, and Fe/H]$=-4$.
We fit in this case 44 line points and 62 continuum points of the observed 
Li profile of the secondary, using the same automated routine as before, but 
fixing the isotopic ratio at $^6$Li/$^7$Li~$ = 0$~\% and the global broadening 
to the instrumental resolution at $v_{\rm BR}= 2.8$~\kms.
We fit in this case 44 line points and 62 continuum points and extract a 
S/N$\sim$149 from the residuals. 
The best fit gives A(Li)$=1.55\pm0.04$~dex and an EW(Li)$=10.9\pm0.8$~m{\AA}.
The global shift of the best fit is a bit large, but still acceptable according to
the errors on the individual RVs of star B. 
This is a downward revision the Li abundance of the secondary  
by $0.22$\,dex relative to the 1D NLTE Li abundance of A(Li)$=1.77$ given in 
\citet{gon08aa} and is consistent with the higher quality data presented in 
the present work.

\section{Discussion and conclusions} \label{sec:con}

The Li abundance of the secondary appears to be significantly lower than that
of the primary, by about 0.6~dex. This result strengthens the increased  
scatter of Li abundances in unevolved metal-poor stars with [Fe/H]$<-3.5$~dex. This
increased scatter and possibly meltdown of the Li plateau at the lowest metallicities
\citep{sbo10aa} is also consistent with the similar Li abundance at roughly the level
of the Li plateau (A(Li)$_{\rm A-B}=0.1$~dex) for both components 
(with $T_{\rm eff,A}=6350$~K and $T_{\rm eff,B}=5830$~K) of the doubled-line 
spectroscopic binary G166--45 at a metallicity [Fe/H]$=-2.5$~\citep{aok12apjl}. 
These authors discussed it in the context of the mass-dependent Li depletion 
suggested by \citet{mel10}, which does not seem to be supported by the Li 
abundances of these two spectroscopic binaries.  

\begin{figure}
\begin{center}
{\includegraphics[clip=true,width=85mm,angle=180]{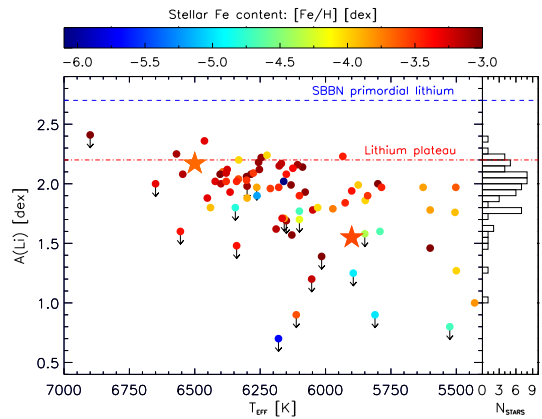}}
\end{center}
\caption{
Lithium abundance vs. effective temperature 
in CS\,22876-032A and B (large  stars), 
compared with the literature measurements (small circles) in unevolved stars 
with [Fe/H]$<-3.0$ and logg$>3.0$. 
Downward arrows indicate  upper limits. 
The histogram on the right side  shows the Li measurements without including 
the upper-limit values. The blue dashed line is the predicted SBBN 
primordial lithium abundance and the red dash-dotted line is the 
value of the lithium plateau, known as the {\it Spite Plateau}.
\label{fig:liteff}
}
\end{figure}

\begin{figure}
\begin{center}
{\includegraphics[clip=true,width=85mm,angle=180]{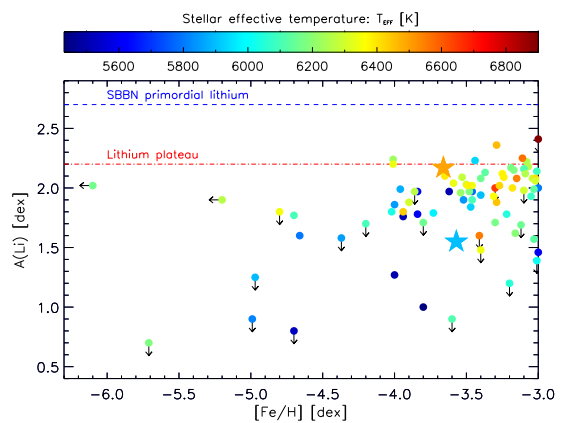}}
\end{center}
\caption{
Lithium abundance vs. metallicity [Fe/H]
in CS\,22876-032A and B (large stars), 
compared with the literature measurements (small circles) in unevolved stars 
with [Fe/H]$<-3.0$ and logg$>3.0$. 
Downward arrows indicate  upper limits. 
The blue dashed line is the predicted SBBN 
primordial lithium abundance and the red dash-dotted line is the 
value of the lithium plateau, known as the  {\it Spite Plateau}.
\label{fig:life}
}
\end{figure}

In Fig.~\ref{fig:liteff} and Fig.~\ref{fig:life} 
we compare the Li abundances of star A and B of CS~22876--032 
with the literature measurements for stars with [Fe/H]$\le -3.0$ and $\logg > 3$, 
showing the meltdown of the lithium plateau~\citep{sbo10aa}. 
In this metallicity regime, most of the stars are located around A(Li)$\sim2.0$ and 
there is an evident lack of stars  between this upper envelope defined by the 
lithium plateau and the predicted SBBN primordial Li abundance. This is clearly seen
if we only consider  the measurements; leaving aside the upper limits, there  
appears to be a trend of Li abundance with effective temperature, together with 
increasing scatter towards lower temperatures. 
This trend is expected as a result of convection being stronger in cooler stars, 
thus resulting in stronger Li depletion. 
What is not expected is that the Li depletion begins at such warm effective 
temperatures at \teff\ $\sim 6100$\,K. 
At metallicities [Fe/H]~$\sim -2$ the same pattern happens at 
\teff\ $\sim 5800$\,K~\citep{bon97mnras}.
Based on this evidence~\citet{bon18aa} have suggested that the Li depletion starts at 
increasingly higher effective temperatures with decreasing metallicity. 
The binary CS 22876--032 provides strong support to this claim since the two stars 
should have  formed with the same Li abundance. 

The Li upper limits complicate the picture since some are found at very high effective 
temperatures. The stars with \teff $> 6500$\,K and no Li detected are   
SDSS~J150702+005152 \citep{bon18aa}, and HE~1029-0546 and HE~1241-2907 \citep{han15apj}.
\citet{ryan01apj} have suggested that the stars with no measurable Li are 
future blue-stragglers  and that the absence of Li in these stars is the result of 
the merging of two main-sequence stars that form the blue straggler.
\citet{boni19rnaas} have shown that three of the four stars studied by \citet{ryan01apj} 
are indeed canonical blue stragglers. This suggests a simple interpretation of 
Fig.~\ref{fig:liteff} and the meltdown of the Spite plateau 
(see also Fig.~\ref{fig:life}): 
all stars with no measurable Li are blue stragglers or blue-stragglers-to-be, 
while all stars with measurable Li were formed with a Li abundance equal to the 
Spite plateau value, and the cooler stars underwent depletion due to convection. 
This hypothesis would be disproved if any of the upper limits at A(Li)$< 1.5$ were 
turned into an actual measurement. A blue straggler has no measurable Li.   

Another way to probe the effects of convection in the stars of this system is to 
measure the Be abundance in the two stars. If convection has been so vigorous to 
substantially deplete Li in CS~22876--032~B, it may also have  been able to deplete 
Be and thus the two stars would also have different Be abundances.
However, regarding Be, the picture might be considerably more complex.
\citet{gon08aa} found an upper limit log(Be/H)$< -13.0$ and we concluded that 
since this was one order of magnitude higher than expected for stars at this 
metallicity it was not significant.
We overlooked the fact that the star is highly enhanced in oxygen and thus we should 
compare this Be upper limit with the Galactic Be-O trend. 
CS~22876-32~A has [O/H]= --1.56 in 1D LTE (we assumed a solar oxygen abundance 8.76 
from \citealt{ste15aa}).
If we consider the straight line fit to the 1D~LTE data of \citet{boe11apj} we find 
that for this oxygen abundance we expect log(Be/H)$=-12.74$. Thus, our upper limit is 
marginally significant, we should have measured Be.
It could simply be that this system is one that significantly deviates from the 
Galactic Be-O relation. \citet{spi19aa} found that 2MASS\,J1808-5104, which
has a similar [Fe/H] but lower [O/H] than CS\,22876-32, has no
detectable Be and its upper limit lies well below the Galactic Be-O line. 
Another similar case is the CEMP-no star BD +44$^\circ$~493 
\citep{ito13apj}, again a similar [Fe/H] and no detectable Be.
Both stars appear in Fig.\,\ref{fig:liteff} at the lowest \teff\ and are Li depleted, 
but with measurable Li.
The explanation suggested by \citet{spi19aa} for these stars was that their age is old 
enough that cosmic rays did not have enough time to build significant quantities of Be.
New higher-quality observations of Be in CS\,22876-32 would be very valuable, 
since they would allow us to establish if this system lies on the Galactic Be-O line or 
if it is below the line, like 2MASS~J1808-5104 and BD +44$^\circ$~493. 

The analysis of the high-quality UVES spectrum of the metal-poor binary CS 22876-032 
using 3D-NLTE synthetic spectra of the \ion{Li}{i} $\lambda$670.8~nm doublet does not 
provide a strong detection of the $^6$Li isotope. 
The results of the primary star can be interpreted as an upper limit  of the Li 
isotopic ratio of the $^6$Li/$^7$Li~$< 10$\%. However, the result is also consistent 
with no detection of $^6$Li. 
This together with the upper limits and/or non-detections of $^6$Li in other 
single metal-poor stars~\citep{ste12msais,lin13aa} when a 3D-NLTE analysis is 
performed seem to solve the so-called {\it \emph{second cosmological Li problem}}, and 
therefore support the negligible amount of $^6$Li predicted in standard Big Bang 
nucleosynthesis models. 
Future observations performed with the ESPRESSO spectrograph at the 8.2m VLT
\citep[e.g.][]{pep14,gon18} using the UHR mode at $R\sim 200,000$ with a pixel size 
of 0.5 km/s and fiber diameter of 0.5 arcsec may help to further investigate the 
lithium isotopic ratio and velocity fields of these and other metal poor stars.

\begin{acknowledgements}
JIGH acknowledges financial support from the Spanish Ministry project 
MINECO AYA2017-86389-P, and also from the Spanish MINECO under the 2013 
Ram\'on y Cajal program MINECO RYC-2013-14875. 
PB and EC acknowledge support from the Scientific Council of Observatoire de 
Paris and from the Programme National de Physique Stellaire of the Institut 
National des Sciences de l'Univers – CNRS.
HGL acknowledges financial support from the Sonderforschungsbereich SFB\,881
``The Milky Way System'' (subproject A4) of the German Research Foundation
(DFG).
\end{acknowledgements}

%
%

\bibliographystyle{aa} 
\bibliography{mpoor} 

\end{document}